# Interface structure, band alignment and built-in potentials at LaFeO$_3$/$n$-SrTiO$_3$ heterojunctions

Ryan Comes[1,2] and Scott Chambers[1]

[1]Physical and Computational Sciences Directorate
Pacific Northwest National Laboratory
Richland, WA 99352

[2]Department of Physics
Auburn University
Auburn, AL 36849

**Abstract**

Interface structure at polar/non-polar interfaces has been shown to be a key factor in controlling emergent behavior in oxide heterostructures, including the LaFeO$_3$/$n$-SrTiO$_3$ system. We demonstrate via high energy resolution x-ray photoemission that epitaxial LaFeO$_3$/$n$-SrTiO$_3$(001) heterojunctions engineered to have opposite interface polarities exhibit very similar band offsets and potential gradients within the LaFeO$_3$ films. However, differences in the potential gradient within the SrTiO$_3$ layer depending on polarity may promote hole diffusion into LaFeO$_3$ for applications in photocatalysis.

Semiconducting perovskite oxides with the formula $AB$O$_3$, where $B$ is a transition metal ion and $A$ is a rare-earth or alkali-earth ion, are of growing interest for photocatalytic water splitting. Oxides are far more stable in the terrestrial atmosphere than are traditional semiconductors. Likewise, the prospects for stability in aqueous solutions are better for oxides, particularly near neutral pH. For good solar photocatalytic performance, a band gap of approximately ~1.5 eV is ideal [1], making Fe-based oxides such as α-Fe$_2$O$_3$ [2] and LaFeO$_3$ [3,4] (LFO) attractive candidates. A 2.3 eV direct bandgap has been reported for epitaxial thin films of LFO [5], while doping with Sr enhances light absorption at lower photon energies [6]. Effective spatial separation of optically-excited electron-hole pairs is also desirable to enhance carrier lifetimes. One approach to solve this problem is the use of a ferroelectric oxide which has an intrinsic electric field to separate electrons and holes [7]. This approach has been used to good effect in BiFeO$_3$ epitaxial thin films with a band gap comparable to LFO [8].

Recent reports of interface induced polarization for LFO films grown on Nb-doped SrTiO$_3$ ($n$-STO) have spurred interest in understanding electronic reconstruction at the junction of these perovskites [9,10]. M. Nakamura *et al.* [9] report that growing LFO on both SrO- and TiO$_2$-terminated $n$-STO(001) substrates results in polarization in the films, but with the polarization dependent on the interface structure. SrO-terminated STO(001) was generated by preparing a TiO$_2$-terminated substrate using conventional wet chemical etching using HF acid [11] followed by the deposition of a single monolayer of SrO. For LFO films grown on SrO-terminated (TiO$_2$-terminated) STO, the idealized heterojunction would be negatively (positively) charged with an SrO$^0$-FeO$_2^-$ (TiO$_2^0$-LaO$^+$) interface. These authors report zero-bias, visible-light photocurrents flowing in opposite directions depending on the interfacial termination. This result was attributed to bulk polarizations in the material, as previously predicted under extremely high strain (~9%) in rare-earth ferrites [12]. In the absence of high strain (the lattice mismatch between LFO and *n*-STO is only ~1%), the authors posit that the induced polarization is driven by differing interface dipoles and polar discontinuities. Subsequently, independent work by K. Nakamura *et al.* [10] employed an identical experimental approach and also found that visible-light photoconductivity at zero bias was dependent on the interface structure. In the work of K. Nakamura [10], visible-light photocatalytic responses were also observed with minimal dependence on interfacial structure for $h\nu < 2.9$ eV. A much stronger response was observed for $h\nu > 3.2$ eV (i.e. above the band gap of STO) for specimens with the SrO-FeO$_2$ interface. This result was attributed to differing electronic reconstructions leading to either accumulation (TiO$_2$-LaO$^+$) or depletion (SrO-FeO$_2^-$) at the heterojunction. These intriguing results motivate additional experiments aimed at detecting the presence (or absence) of oppositely oriented built-in potentials and polarizations within LFO films.

To this end, we prepared a series of LFO/*n*-STO(001) heterojunctions using oxygen-assisted molecular beam epitaxy (MBE) and carried out interface electronic structure measurements using high energy resolution x-ray photoelectron spectroscopy (XPS) and spectroscopic ellipsometry (SE). All films were grown at 600 ± 50°C at a rate of one LaO or FeO$_2$ monolayer every 43 seconds using effusion cells and alternately shuttering the La and Fe beams, with a mixed O/O$_2$ beam generated by an electron cyclotron resonance source continuously incident on the substrate [13]. A pair of 0.05% Nb-doped STO substrates (Crystec) were prepared side-by-side using a boiling deionized water treatment [14], followed by an anneal in air at 1000 °C for 30 minutes. The samples were then cleaned in ozone on the bench and loaded into an oxide MBE system (DCA) with an appended x-ray photoelectron spectrometer (VG Scienta R3000 analyzer and monochromatic Al Kα x-ray source). The TiO$_2$ surface termination was confirmed using angle-resolved XPS measurements [15]. A single monolayer of SrO was then deposited using an effusion cell on one of the substrates to achieve the A-site termination, also confirmed by angle-resolved XPS. After preparing the two substrates, increments of three unit cells (u.c., 1 u.c. = ~3.9Å) of LFO were grown with a shuttering sequence configured to match the substrate termination (i.e. FeO$_2$ (LaO) layer deposited on the SrO- (TiO$_2$-) termination). *In situ* reflection high-energy electron diffraction (RHEED) was used to monitor the surface structure during growth, as can be seen in Figure S1 [15].

*In situ* XPS band alignment measurements were made following each 3 u.c. deposition, with a separate thick (~12 nm) LFO film grown to determine the bulk electronic structure of the material. Details of band alignment analysis are described elsewhere [13] and can be found in the supplemental information [15]. In brief, the energy separation between core levels unique to the film and the substrate were used to extract the VBO, assuming that there is no differential charge accumulation between the film and substrate due to photoemission. We saw no evidence of charging for the 3, 6 and 9 u.c. films. However, the 12 nm LFO films did accumulate charge and a low-energy electron flood gun was used to neutralize this charge. The energy scale was calibrated using a constant offset such that the O 1$s$ singlet was at the same binding energy as that for the 9 u.c. *A*-site terminated LFO film (529.2 eV).

Figure 1 summarizes the relevant spectra for all LFO/*n*-STO heterostructures, STO substrates, and thick LFO films. The Sr 3$d$ and La 4$d$ peaks Figure 1(a) have been shifted slightly so that all Sr 3$d_{5/2}$ peaks are aligned at 0 eV as a means of visualizing the separation between the Sr 3$d$ and La 4$d$ peaks, which increases with increasing thickness. This monotonic shift reveals changes in built-in potential and band alignment with thickness, as has been reported previously in LaCrO$_3$ films grown on STO [17,18], and in LFO films grown on arbitrarily terminated *n*-STO [19]. The La 4$d$ peaks for both *A*- and *B*-site terminated films move to lower binding energy with increasing thickness, indicating that the VBO increases with thickness, and that the LFO valence band maximum (VBM) moves closer to the Fermi level ($E_F$). From these spectra, the valence band offsets (VBO) are readily determined, as described below.

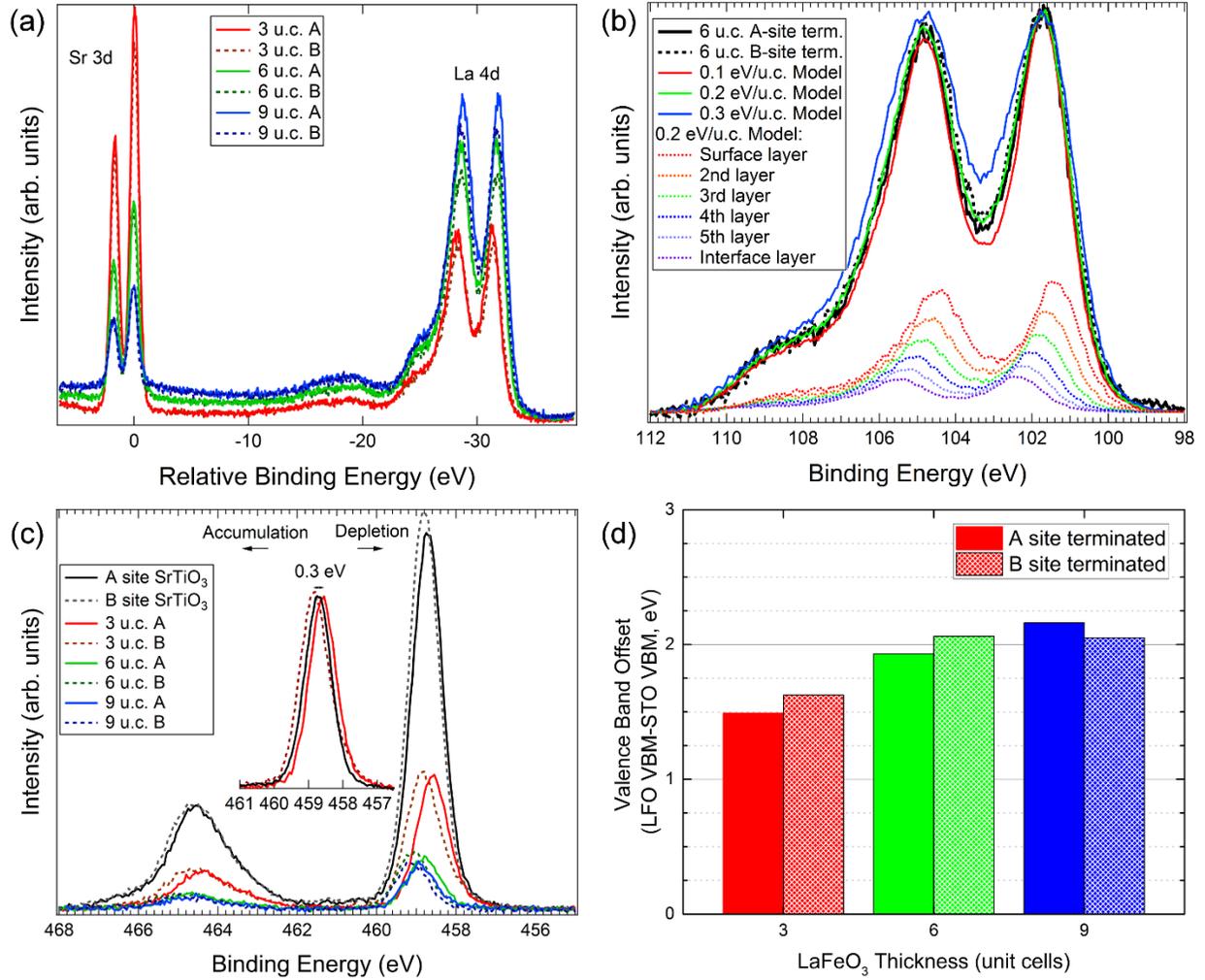

**Figure 1: (a)** Sr 3d and La 4d core-level spectra for the family of heterostructures, shifted to align the Sr 3d peaks; **(b)** Model of La 4d peak broadening in the 6 u.c. films; **(c)** Ti 2p core -level spectra for each film and substrate, with the inset showing the peak shifts; **(d)** Valence band offsets determined from the core-level spectra for each heterojunction.

The La 4$d$ core level peaks for the thin-film heterostructures are broader than those for the thick reference LFO films, which are nominally flat band. This result is consistent with the presence of a built-in potential within the thinner films. By modeling this broadening to account for contributions of each unit cell, we can estimate these potential gradients [15]. The broadening is nearly identical for both the *A*-site and *B*-site terminated samples for each thickness, revealing that the potential gradient within the LFO is the same for both interface terminations. This experimental result contradicts previous work in which potential gradients of opposite sign were assumed for the two terminations in order to explain photoconductivity data [9]. We measure built-in potentials of 0.3 eV/u.c. for the 3 u.c. film and 0.2 eV/u.c. for the 6 u.c. film for both interface polarities. This allows us to estimate the change in electrostatic potential from the film surface to the film-substrate interface. The uncertainty in the potential in each layer is estimated by varying the potential at each unit cell in the stack until the fit to the data is clearly worse than the initial model. This produces uncertainties of 0.1 eV near the film surface and 0.2-0.3 eV deeper within the film. We see comparable broadening between the 6 and 9 u.c. thick films, but our analysis indicates that the 9 u.c. film has reached a flat-band condition after completion of 6 u.c.

Although there is no measurable dependence of the potential gradient within the LFO on interface structure, more subtle observations help to elucidate the nature of the interface. As can be seen in Figure 1(c), the Ti 2$p$ core-level binding energy and peak shape depend on the interfacial termination. The TiO$_2$-terminated heterojunctions exhibit uniformly higher binding energies than their SrO-terminated counterparts. Furthermore, the Ti 2$p_{3/2}$ peak for the TiO$_2$-terminated 3 u.c. heterojunction exhibits an asymmetry to low binding energy, in contrast to the STO substrates and the SrO-terminated 3 u.c. interface, all of which are symmetric (see inset). Similar behavior is seen in the 6 u.c. TiO$_2$-terminated case, but the strong attenuation from the thicker 9 u.c. film prevents rigorous analysis for the 9 u.c. sample. No broadening is observed for any SrO-terminated sample. The Ti 2p$_{3/2}$ peak energy for the *B*-site terminated interface is shifted 0.3 eV to higher binding energy compared to the *A*-site terminated sample. This difference is consistent with the presence of fixed positive (negative) charge at the TiO$_2^0$-LaO$^+$ (LaO$^+$-FeO$_2^-$) interface. Similar binding energy shifts are observed for the Sr 3$d$ core level. The low binding energy shoulder seen in the Ti 2$p_{3/2}$ spectra for the B-site terminated specimens could result from outdiffused Ti within the first u.c. of LFO. The energy difference between this feature and the substrate Ti 2$p_{3/2}$ peak is consistent with the valence band discontinuity when a correction is made for the width of the Fe 3$d$ derived feature at the top of the LFO VB, as seen in Fig. 3(a) and described in the Supplemental Information [15]. The *B*-terminated interface between STO and LaAlO$_3$ is also known to exhibit a greater amount of intermixing in previous work, [20] in agreement with our conclusion.

The valence band offsets (VBO) determined from fits of the Ti 2$p$, Sr 3$d$, La 4$d$, and Fe 3$p$ spectra in Figure 1(a-c) are shown in Figure 1(d). The uncertainties in these measurements are ±0.1 eV in each case. We find that the interfacial termination has a negligible effect on the VBO, indicating that there is no divergence in the potential due to an induced ferroelectric-type polarization, in contrast to the behavior others have reported [9,10]. Instead the VBO increases with increasing LFO thickness and saturates at 2.0-2.2 eV, close to the value previously reported for arbitrarily-terminated *n*-STO [19].

To estimate the conduction band offset, SE measurements were performed to determine the band gap of LFO using a 12 nm thick film grown on undoped STO. The results of fits to the data are shown in Figure 2. Various groups have used values of 2.0 [19], 2.1 [10], and 2.2 eV [9] in their analysis of the LFO/*n*-STO band alignment. However, computational and experimental studies by Scafetta, *et al*. [5] have shown that the band gap is 2.3 eV when the optical transition is modeled as a direct--forbidden excitation from the hybridized majority-spin Fe 3$d$ $e_g$ and O 2$p$ VB maximum to the unfilled minority-spin Fe 3$d$ $t_{2g}$ conduction band (CB) minimum. This method involves extrapolating the linear region of the quantity $(\alpha h\nu)^{2/3}$ to the energy axis. Here $\alpha$ is the absorption coefficient and *hv* is the photon energy. Using the same approach (see Figure 2), we find an identical band gap of 2.3 eV.

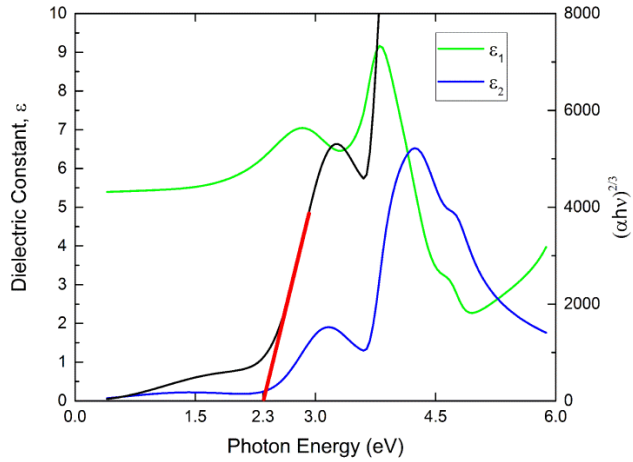

**Figure 2: Spectroscopic ellipsometry measurements of dielectric constants for a 12 nm LFO film on STO(001) (left axis) and determination of direct forbidden band gap (black/red, right axis).**

Based on our measurements of the VBO and the LFO band gap, along with the known STO indirect gap of 3.25 eV [16], we can construct energy diagrams for the LFO/*n*-STO heterostructure as a function of thickness. These are shown in Figure 3, along with XPS valence band measurements for each sample. The points at -10 u.c. in Figure 3(b-d) represent the bulk and were calculated assuming an effective density of states at the conduction band edge for the distribution of electrons in Nb:STO. The LFO bands were determined based on the measured VBO and built-in potential gradients determined in Figure 1, with the CBO determined using the LFO gap determined from ellipsometry [15]. The valence band maxima for the various LFO films extracted from the spectra shown in Figure 3(a) are in excellent agreement with the values extracted from the energy diagrams shown in Figure 3(b-d).

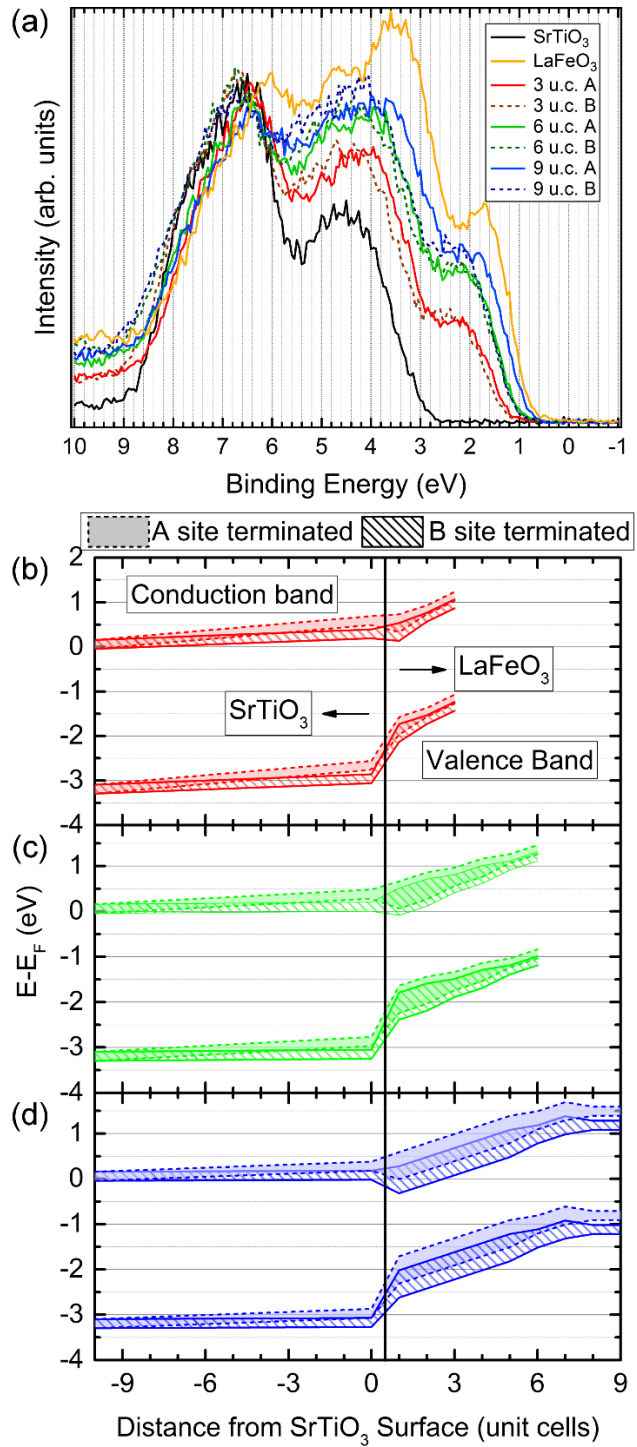

**Figure 3: (a) Valence band XPS spectra for bulk STO(001), a thick film of LFO(001), and all LFO/STO(001) heterojunctions (a). VBM (solid squares) and CBM (open squares) values taken from core-level binding energies for (b) 3, (c) 6, and (d) 9 u.c.LFO/n-STO. The values at 10 u.c. are representative of bulk STO and were calculated as described in the text.**

Others have concluded that LFO is degenerately *p*-type and that the interfacial conduction band offset (CBO) is ~1 eV [19]. However, we find that the LFO and STO conduction band minima are nearly degenerate at the interface for all thicknesses investigated. This result indicates that LFO is *n*-type at the interface, with the possibility of some $Fe^{2+}$ character due to carrier leakage from the *n*-STO. This finding is in agreement with other reports which have shown that electron transfer occurs at the LaTiO$_3$/LFO interface, resulting in $Fe^{2+}$ [21], and that $Fe^{2+}$ is present in co-doped La$_{2x}$Sr$_{1-2x}$Fe$_x$Ti$_{1-x}$O$_3$ in which two La donors are present for every Fe acceptor [22]. After 6-9 unit cells, however, the conduction band levels off at higher energy and $E_F$ is approximately mid-gap in the LFO, making it an intrinsic semiconductor. This can be understood when one considers the instability both of holes from $Fe^{4+}$ in (La,Sr)FeO$_3$ [6] and electrons from $Fe^{2+}$ in Fe$_3$O$_4$ [23] in ambient conditions, indicating the high stability of $Fe^{3+}$.

The energy differences between STO VBM values for interfaces prepared from A- and B-terminated STO are consistent with the presence of fixed charge of opposite sign. The VBM is closer to the Fermi level by 0.2 – 0.3 eV when the interface structure is nominally $(SrO)^0$ – $(FeO_2)^{1-}$ than it is for the $(TiO_2)^0$ – $(LaO)^{1+}$ interface structure. We say "nominal" because cation mixing occurs to some extent, resulting in modified interface charges. The difference in STO valence band energy between the two heterojunctions reveals that there is a difference in the magnitude of the potential gradient in the STO for the two heterostructures. The energy diagrams we present in Figures 3(b-d) reveal that the potential on the STO side of the $(TiO_2)^0$ – $(LaO)^{1+}$ interface is nearly constant, while there is upward band bending on the order of 0.2-0.3 eV for the $(SrO)^0$ – $(FeO_2)^{1-}$ case. The latter should promote hole drift from STO towards the LFO film. This result anecdotally supports the observations of K. Nakamura *et al*. [10], who rationalized variations in photocatalytic response by assuming downward band bending in the STO near the $(TiO_2)^0$ – $(LaO)^{1+}$ interface, which, our data do not support. Qualitatively similar behavior might be observed if, instead, there is upward band bending that promotes hole drift towards the surface near the $(SrO)^0$ – $(FeO_2)^{1-}$ interface and a flat-band condition at the $(TiO_2)^0$ – $(LaO)^{1+}$ interface. In general, however, the band alignments that we observe are not expected to produce the dramatic differences in photoresponse reported in Refs. 9 and 10. Our results do not support the hypothesis of polarization in LFO pointing in opposite directions and we suggest that other as yet undetermined mechanisms may be operative.

In summary, we synthesized a series of epitaxial LaFeO$_3$ films on *n*-doped SrTiO$_3$ with opposite surface terminations to study the effect of interface polarity on band alignment, potential gradients, and electronic reconstruction at the interface. We find that interface polarity has a negligible effect on the potential gradient within the films, in disagreement with previous reports of interface-controlled polarization in LFO. We do measure a slight difference in band bending on the STO side of the interface, presumably resulting from screened fixed charge of opposite sign for the two interface polarities. However, it seems unlikely that this difference could account for the substantial differences in photoresponse reported in recent literature. We also have found the conduction band alignment between *n*-STO and LFO, showing that the CBO is ~0 eV at the interface. These results provide important insights to understand the LFO/*n*-STO heterojunction for solar energy applications.

RBC was supported by the Linus Pauling Distinguished Post-doctoral Fellowship at Pacific Northwest National Laboratory (PNNL LDRD PN13100/2581). SAC was supported by the U.S. Department of Energy (DOE), Basic Energy Sciences (BES), Division of Materials Sciences and Engineering under Award #10122. A portion of this research was performed using EMSL, a national scientific user facility sponsored by the Department of Energy's Office of Biological and Environmental Research and located at Pacific Northwest National Laboratory.

# Supplemental Information for

# Interface structure, band alignment and built-in potentials at LaFeO$_3$/$n$-SrTiO$_3$ heterojunctions

Ryan Comes[1,2] and Scott Chambers[1]

[1]Physical and Computational Sciences Directorate
Pacific Northwest National Laboratory
Richland, WA 99352

[2]Department of Physics
Auburn University
Auburn, AL 36849

**Film Synthesis**

Film growth was monitored *in situ* using reflection high-energy electron diffraction (RHEED). As we have observed previously [1], the RHEED pattern is more streaky for the *A*O terminated sample, while it exhibits sharper spots for the *B*O$_2$ terminated case. This is shown in Figure S1.

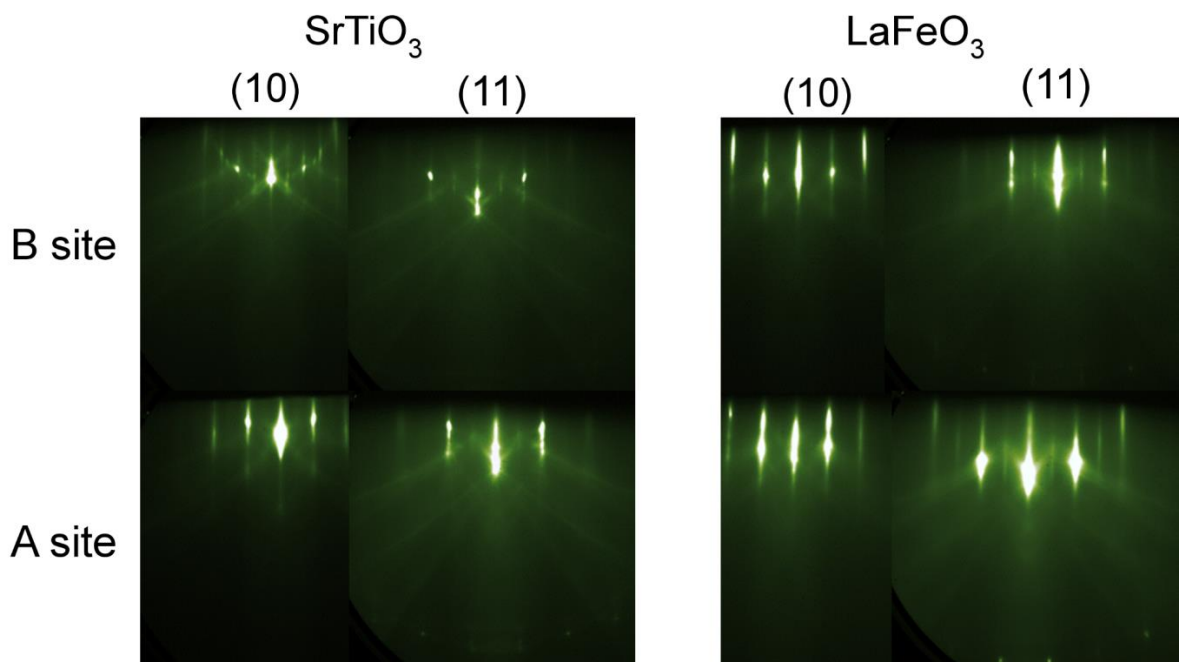

Figure S4: Representative RHEED patterns for SrTiO$_3$(001) substrates and 6 u.c. LaFeO$_3$ films with specified terminations.

**Valence Band Offset Analysis**

The use of core-level x-ray photoelectron spectroscopy (XPS) is well established for the determination of valence band alignment at a buried interface. Reference spectra for bulk or bulk like materials are required to perform the analysis. These can be in the form of either a single crystal, such as $n$-doped SrTiO$_3$ (STO), or a thick epitaxial film, as in the case of LaFeO$_3$ (LFO). By fitting various core electron peaks and determining the valence band maximum (VBM), the differences in binding energy between the cores and the VBM are established for each material. These quantities (labeled $\Delta E_{Ref}$ in the text and equations below) then allow the VB maxima for the two materials in the heterojunction to be determined from their core-level binding energies. This analysis is accurate provided there are no significant changes in core-level line shape for the heterojunction as a result of interface chemistry. In this work, we used the Ti $2p_{3/2}$, Sr $3d_{5/2}$, and Ti $3p$ peaks for STO, and the La $4d_{5/2}$ and Fe $3p$ peaks for LFO. For each heterostructure, fits to each of these peaks were used to determine the VBM for the buried STO substrate and the LFO film using the formula

$$E_{VBM} = E_{CL} - \Delta E_{Ref} \qquad (1)$$

Here $E_{CL}$ represents the core-level binding energies measured for each material. This approach allows us to establish the VBM of each material using multiple pairs of core peaks to reduce uncertainty. The variation in VBM from different core levels in a given material (STO or LFO) was ~0.1 eV, establishing the uncertainty for these measurements. To determine the valence band offset (VBO), the difference between the VBM values for LFO and STO was determined, leading to,

$$E_{VBO} = E_{VBM,STO} - E_{VBM,LFO} \qquad (2)$$

This approach is equivalent to that based on the separation between core-level peaks for the two materials [1], such as Sr $3d_{5/2}$ and La $4d_{5/2}$. As noted in the Letter, the difference between the Sr and La core levels varies in a manner consistent with a changing VBO.

We have shown that there are potential gradients both in the near-interface STO layers for A-type interfaces, and in all LFO films. Accordingly, the most precise determination of the VBO would in principle require using spectra that are representative of the layers *directly* at the interface, as obtained from the modeling (e.g. see Figure 1b in the Letter). However, doing so would add a new and unknown level of uncertainty resulting from extracting layer-specific binding energies from the model. Therefore, we chose a simpler approach using the broadened spectra as measured, understanding that the resulting VBOs are average values over XPS probe depth. The uncertainty bands in Figure 3 of the Letter reflect this reality.

**Surface Termination**

To determine the surface termination of the substrates prior to growth, angle-resolved XPS was employed. Measurements of the Sr 3$d$ and Ti 2$p$ at normal emission (0°) and 70° off-normal allow the probe depth to be varied, as shown by the formula

$$I(\theta, d) = I(\theta, 0)e^{-d/\lambda \cos \theta} \quad (3)$$

Here, $I$ is the photoelectron intensity from a given layer at a depth $d$ below the surface, $\theta$ is the angle relative to the surface normal, and $\lambda$ is the electron attenuation length [3]. Thus, the effective probe depth decreases as $\theta$ increases, and the intensity becomes more sensitive to the surface termination. In practice, intensities measured at large $\theta$ are also reduced due to geometric factors, but the overall effect remains extremely useful. To verify that the primary termination for each substrate matched our expectations, we measured the ratio of the Sr and Ti core-level peak areas at both angles and compared the results, as seen in Table S1. Variations between STO1 and STO2 prior to SrO monolayer deposition at 70° are most likely attributable to photoelectron diffraction effects due to different azimuthal orientations of the two samples, since the normal emission ratios are nearly identical. Nevertheless, the changes in Ti 2$p$/Sr 3$d$ peak area ratios resulting from the SrO monolayer deposition qualitatively reveal a change in surface termination, as expected.

**Table S1: XPS Peak Intensities for Substrates Prior to Growth**

|  | STO1 for *A*-term before SrO |  | STO2 for *B*-term |  | STO1 after SrO |  |
|---|---|---|---|---|---|---|
| **Angle (°)** | 0 | 70 | 0 | 70 | 0 | 70 |
| **Ti 2p** | 579412 | 274478 | 580508 | 312796 | 544114 | 146053 |
| **Sr 3d** | 479840 | 211290 | 483389 | 184257 | 548665 | 179557 |
| **Ti 2p/Sr 3d** | 1.21 | 1.30 | 1.20 | 1.70 | 0.99 | 0.81 |

**Cation Intermixing**

To probe cation intermixing at the STO/LFO interface using XPS, we must determine accurate attenuation lengths specifically for LFO. To this end, we measure peak areas for the Fe 2$p$ and La 4$d$ core levels for a range of thicknesses. As the film thickness increases and exceeds the electron attenuation length by a considerable margin, the peak areas should approach asymptotes once the measurement increasingly probes only the LFO film and none of the underlying STO. The La and Fe intensities measured at normal emission ($\theta = 0°$), $I$, at normal emission can be modeled as

$$I(t) = I(\infty)(1 - e^{-\frac{t}{\lambda}}) \quad (4)$$

Here, $I(\infty)$ is the intensity from a suitably thick reference LFO film, $\lambda$ is the attenuation length and $t$ is the film thickness. By fitting this equation to Fe 2$p$ and La 4$d$ peak areas in the *A*- and *B*- terminated films of different thicknesses, we can determine $\lambda$ at the two binding energies. These data and the associated fits are shown in Figure S5. We find excellent agreement for the fits between our two samples, with $\lambda_{Fe2p}$ equal to 9.7(8) Å and $\lambda_{La4d}$ equal to 15.5(9) Å. Assuming that the attenuation length for a given photoelectron kinetic energy $E_k$ will scale as $\lambda \propto E_k^n$, where $n$ is an exponent less than one, the resulting values of $n_{Fe2p}$ and $n_{La3d}$ are 0.34(1) and 0.38(1).

These results can be used to interpret Ti 2$p$ and Sr 3$d$ core-level intensities from the STO substrate and determine the extent of intermixing across the interface. Using these exponents for Fe 2$p$ and La 4$d$ photoelectrons traversing through LFO, we estimate $\lambda$ to be 15(1) Å and 12(2) Å for the Sr 3$d$ and Ti 2$p$ peaks, respectively. Here we have estimated $n$ to be 0.36(2) for Ti 2$p$, since the kinetic energy of Ti 2$p$ is nearly half way between the values for La 4$d$ and Fe 2$p$.

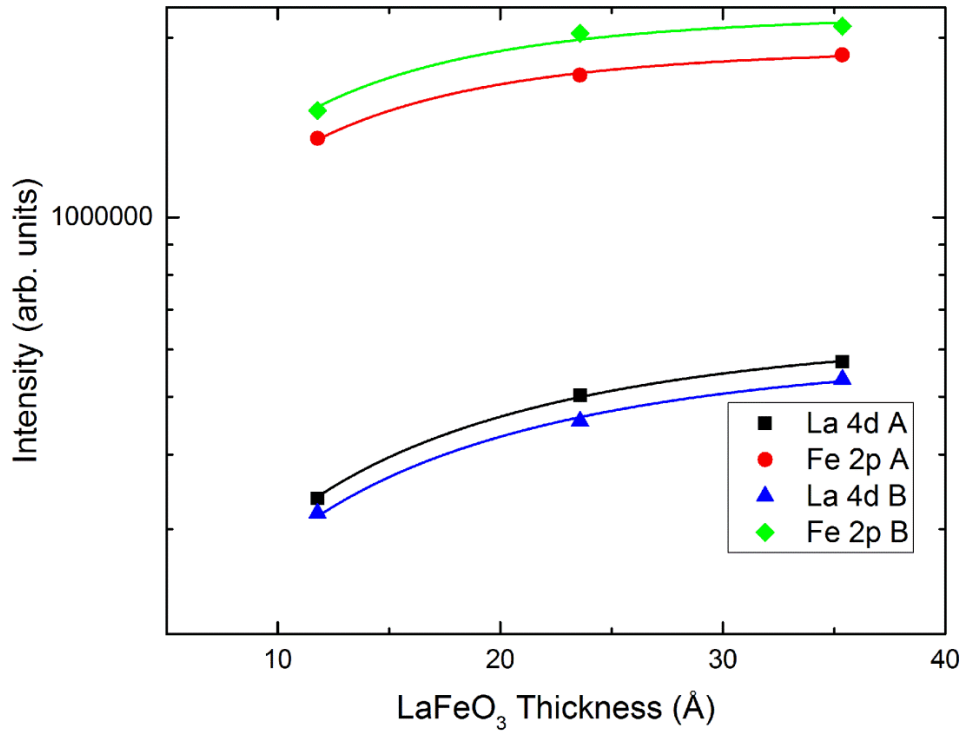

**Figure S5:** Peak areas and model fits for the La 4$d$ and Fe 2$p$ core levels in epitaxial LFO films of thickness 3, 6 and 9 u.c. grown on STO(001).

To assess the extent of cation mixing at the interface, we consider the rate of decay of the Sr 3$d$ and Ti 2$p$ intensities measured at normal emission as a function of LFO film thickness. We model the intensity $I$ as

$$I(t) = I_0 e^{-t/\lambda}. \tag{5}$$

Here, $I_0$ is the core electron intensity for clean STO(001), $t$ is the thickness of the LFO film, and $\lambda$ is the attenuation length. If the interface is atomically abrupt, the attenuation lengths we extract from the fits of eqn. 5 to the data should match those resulting from the fits of the data to eqn. 4. If there is significant outdiffusion of Ti and/or Sr into the LFO film, the attenuation lengths for Ti 2$p$ and Sr 3$d$ from fitting measured peak areas to eqn. 5 would be larger than those from fitting to eqn. 4.

The fits to eqn. 5 are shown in Figure S6. From these fits, we extract an attenuation length of 16(1) Å for the Sr 3$d$ peaks and 12.0(5) Å for the Ti 2$p$ peaks for the entire film set. These values are in excellent agreement with those resulting from fits to eqn. 4. We thus conclude that the extent of interfacial cation mixing is slight and, thus, not a major contributor to the results we observe. The presence of a low binding energy shoulder in the Ti 2$p_{3/2}$ peak for 6 u.c. of LFO deposited on B-terminated STO (see Fig. 1c in the Letter) is most likely due to Ti cations that have diffused into the LFO film. However, the results discussed above reveal that Ti outdiffusion is not extensive. Therefore, we conclude that the outdiffused Ti cations which give rise to this shoulder are found only in the first u.c. of LFO.

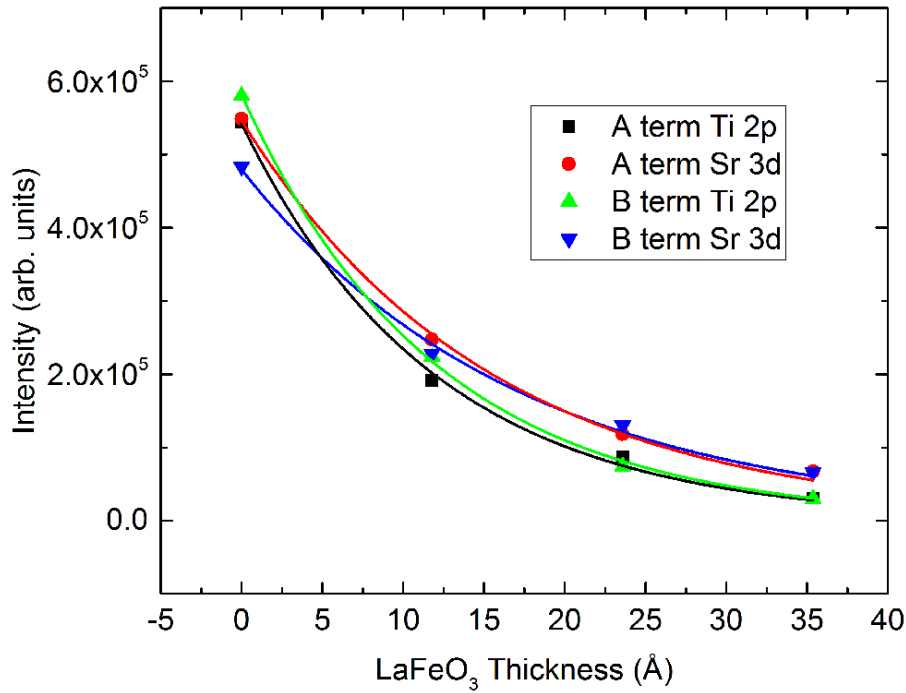

Figure S6: Peak areas and model fits for the Sr 3$d$ and Ti 2$p$ core levels for epitaxial LFO films of thickness 3, 6 and 9 u.c. grown on STO(001).

**Estimation of Potential Gradient**

By modeling the broadening of the complex La 4$d$ peak manifold to account for the effects that the built-in potential gradient have on the measured spectra, we were able to estimate the potential at each layer in the film. This analysis was performed by modeling the contribution from each layer using a flat-band reference spectrum from the LFO thick film and then summing over layers, with the intensities scaled to account for the depths of the various layers, and the peaks shifted to simulate the potential gradient across the film. The fitting function is thus given by

$$I_{model}(E) = \sum_{n=0}^{N-1} I_{ref}(E + n\Delta E)e^{-nc/\lambda} \tag{6}$$

Here, $I_{model}$ is the model spectrum for the heterojunction, $I_{ref}$ is the reference thick-film LFO spectrum, $E$ is the binding energy, $\Delta E$ is the energy change per unit cell relative to the surface unit cell (i.e. 0.1, 0.2. or 0.3 eV/u.c. in Figure 1(b)), $N$ is the number of unit cells included in the model, $c$ is the out-of-plane lattice parameter for the LFO film (taken to be 3.93 Å), and $\lambda$ is the La 4$d$ attenuation length appropriate for LFO (15.5(9) Å). We determine the $\Delta E$ that yields the best fit to the measured spectrum. To determine the uncertainty in the potential at each unit cell, we adjust the model by changing the quantity $n\Delta E$ for that particular unit cell until the quality of the fit is noticeably worse than the simple model. This conservatively generates the widths of the error bands for Figs. 4(b) and 4(c) in Letter.